# Techniques to measure quantum criticality in cold atoms

Kaden R. A. Hazzard[1] & Erich J. Mueller[1]

[1]*Laboratory of Atomic and Solid State Physics, Cornell University, Clark Hall, Ithaca, NY 14850, USA*

**Attempts to understand zero temperature phase transitions have forced physicists to consider a regime where the standard paradigms of condensed matter physics break down[1–4]. These quantum critical systems lack a simple description in terms of weakly interacting quasiparticles, but over the past 20 years physicists have gained deep insights into their properties. Most dramatically, theory predicts that universal scaling relationships describe their finite temperature thermodynamics up to remarkably high temperatures. Unfortunately, these universal functions are hard to calculate: for example there are no reliable general techniques[4,5] to calculate the scaling functions for dynamics. Viewing a cold atom experiment as a quantum simulator[6], we show how to extract universal scaling functions from (non-universal) atomic density profiles or spectroscopic measurements. Such experiments can resolve important open questions about the Mott-Metal crossover[7,8] and the dynamics of the finite density $O(2)$ rotor model[1,9], with direct impact on theories of, for example, high temperature superconducting cuprates[10,11], heavy fermion materials[12], and graphene[13].**

Although our ideas are general, much of our discussion will focus on cold bosonic atoms, such as $^{87}$Rb, trapped in optical lattices. Current experiments on this system are capable of observing quantum critical phenomena. This system displays multiple quantum phase transitions, which reside in distinct and non-trivial universality classes[1]. We will also discuss a number of other experimental realizations with richer quantum critical physics. Our protocols can be applied to any quantum phase transition.

Ultracold bosons in optical lattices are described by the Bose-Hubbard model[6]. Figure 1a illustrates this model's zero-temperature phase diagram with a superfluid phase when the tunneling matrix element $t$ is large compared to the on-site interaction $U$, and a Mott insulating phase in the opposite limit. The superfluid phase is characterized by dissipationless mass transport, analogous to the dissipationless charge transport in a superconductor. It supports arbitrarily low-energy excitations and has an order parameter $\psi$ that vanishes at the phase transition. The Mott insulating phase (lobes shown in Fig. 1a) can be caricatured as having a fixed integer number of particles per site, $n$, set by the chemical potential $\mu$. The Mott insulating state has a excitation gap $\Delta$, which vanishes at the transition.

The Mott insulating phase is strictly defined only at zero temperature, but a phase transition between a superfluid and a normal fluid persists to finite temperature (Fig. 1a). Figure 1b illustrates a slice through the finite temperature phase diagram. This figure shows that the normal fluid is divided into three qualitatively distinct regions separated by smooth crossovers illustrated by dashed lines. At large $t/U$, the superfluid (SF) can be heated through a classical phase transition to a normal fluid (NF) with properties similar to that of a weakly interacting Bose gas. At small $t/U$ the finite temperature properties of the normal fluid (MI) are determined by the zero temperature Mott insulator's gap. A quantum critical region intervenes between these two normal fluids. Slices at fixed $t/U$ look identical to slices at fixed $\mu/U$. In fact, this structure is typical of all second-order quantum phase transitions: the disordered phase generically divides into these three regions[1].

Sufficiently near the quantum phase transition (the shaded region in Fig. 1b), the thermodynamic functions obey scaling relationships. This scaling is typically controlled by two energy scales that vanish at the quantum critical point[1]. One scale is the temperature $T$, while the other depends on the phase: it is the zero temperature superfluid stiffness $\rho_s$ in the superfluid and the gap to single particle excitations $\Delta$ in the Mott phase. Observables, such as the density, can be written as the sum of an non-universal part $n_0$ plus a function $n_u$ that obeys the scaling form

$$n_u(T, \Delta) = T^{d/z} \Psi_n(\Delta/T). \tag{1}$$

Here $d$ is the dimension of the system and $\Psi_n$ is a universal scaling function. Below an upper critical dimension $d_c$, the dynamical exponent $z$ and the universal function $\Psi_n$ are the same for a wide range of models and therefore define universality classes: here $d_c = 4 - z$, and $z$ will be 1 or 2 depending on what part of the Mott lobe one is traversing. The non-universal contribution, $n_0(\mu, U, T)$, is the density of the system for $t = 0$, and as explained in the Supplementary Information is easily calculated or measured in independent experiments. Although atypical, with fine tuning it is possible to have situations where additional dimensionless parameters enter. The finite density $O(2)$ rotor model discussed below is one example.

Here we introduce an experimental protocol to observe this universality and measure key properties such as the dynamic critical exponent. Our technique relies upon the inhomogeneity of cold atoms experiments. Although typical observables in cold atoms, such as the density profile, are non-universal, we show that with appropriate analysis the absorption images of these trapped inhomogeneous clouds reveal the universal properties of corresponding homogeneous systems. We find an additional remarkable result: universal quantum critical scaling persists to much smaller clouds than the features associated with the finite temperature *classical* transitions (Fig. 1a, solid line), which have been studied in Refs.[14–17].

We will view trapped atomic systems as locally satisfying Eq. (1), with a gap $\Delta$ that varies as a function of position in space. As discussed below, our analytic and numerical results in trapped systems validate this Thomas-Fermi approximation. Direct measurement of $\Delta(\mathbf{r})$ is difficult in cold atoms. Instead, we will relate the gap to independently measurable thermodynamic quantities.

We illustrate our procedure by taking the compressibility $\kappa \equiv \partial n/\partial \mu$ as the independently measurable quantity. This is appealing because $\kappa$ can be extracted directly from the density profiles via $\kappa = -(1/m\omega^2 r)\partial n/\partial r$ or from $n$'s shot-to-shot fluctuations[18]. Like the density, the compressibility is the sum of an easily measured non-universal part and a universal part, $\kappa = \kappa_0 + \kappa_u$, with

$$\kappa_u(T, \Delta) = T^{d/z-1}\Psi_\kappa(\Delta/T). \tag{2}$$

Equation (1) is an implicit equation for $\Delta$ which together with Eq. (2) implies that

$$\kappa_u T^{1-d/z} = \Upsilon(n_u T^{-d/z}) \tag{3}$$

where $\Upsilon = \Psi_\kappa \circ \Psi_n^{-1}$ is a universal function. Therefore, in the quantum critical regime a graph of $\kappa_u T^{1-d/z}$ versus $n_u T^{-d/z}$ collapses onto a single curve independent of $\mu$, $t$, $U$, $T$, and any other microscopic couplings.

Figure 2b shows this plot extracted from the density profiles (Fig. 2a) of the $t/U \to 0$ trapped one-dimensional Bose-Hubbard model for $N = 70$ particles. These density profiles are efficiently calculated by mapping the system onto non-interacting fermions (see Supplementary Information.) The figure not only demonstrates the universal collapse in the quantum critical regime, but also validates the Thomas-Fermi approximation. The success of the Thomas-Fermi approximation is due to the fact that as one increases temperature above the quantum critical point the coherence length $\xi$ shrinks, and throughout the quantum critical regime it is typically much smaller than the cloud size. As one approaches the classical phase transition, $\xi$ diverges. Hence even though our technique quite readily detects quantum criticality, signatures of the classical transition may be much harder to see[17].

As shown in Fig. 2b, collapse occurs in a region around the quantum critical point, but deviates at sufficiently large densities and temperatures. In this case, the data collapses within $10\%$ for a density range of roughly $0.9 < n < 1.1$ and temperatures range of $T \lesssim t/4$. This is a generic feature of quantum criticality: the physics is universal only near the quantum critical point, with the size of the universal region determined by microscopic details.

The universal scaling functions in Fig. 2 correspond to those of the dilute Bose gas universality class (the same universality class which governs the edge of all cold bosonic clouds). Although this universality class is quantitatively understood[1,19], at larger $t/U$ (near the tip of the lobe) the Bose-Hubbard model falls into the less trivial $O(2)$ universality class, for which open questions remain. In particular, the quantitative structure of the dynamic scaling functions is unknown and no general methods exist to compute these. Furthermore, there is a more intricate "finite density $O(2)$" universal structure governing the crossover between the $O(2)$ and dilute Bose gas universality classes, about which even less is known[9,13]. We give a simple picture of this physics and its relevance to the Bose-Hubbard model in the Supplementary Information.

Figure 3 shows that in addition to measuring universal scaling functions and the size of the universal region, our method enables one to distinguish the universality classes of the two-dimensional square lattice Bose-Hubbard model using only the density profiles. We calculate the equation of state $n(\mu)$ for a set of $\mu$'s realized in a typical cold atoms experiment, employing the numerically exact worm algorithm quantum Monte Carlo as implemented in the ALPS package[20]. To mimic the effect of the trapping potential, we calculated each density for a finite size $10 \times 10$ system. This length scale corresponds to a trapping potential variation of $\sim 1\%$ along the radial direction under typical conditions. We calculate $\kappa(\mu)$ using the lowest order finite difference derivative on this data. The stochastic error in the quantum Monte Carlo yields results with some noise ($\sim 0.1\%$), similar to that found applying finite differences to real experimental data ($\sim 1\%$ noise in radially averaged density profiles, e.g. in Ref.[18]).

We are able to distinguish these universality classes because they are characterized by different dynamical critical exponents ($z = 2$ for the dilute Bose gas and $z = 1$ for the $O(2)$ model)[1]. One sees collapse only when the correct critical exponent is used. Like Fig. 2, the collapse occurs only sufficiently close to the critical point. The behavior persists up to temperatures on the order of $\gtrsim t$ and for a density range of roughly $0.9 < n < 1.1$ Since we are at the upper critical dimension of the dilute Bose gas model, we expect logarithmic corrections to scaling that are hard to identify on these plots. Consistent with our discussion of the Thomas-Fermi approximation's validity, there is no clear signature of the classical phase transition in these graphs.

Applying this method to Fermi lattice systems[21,22] would allow one to resolve a number of important open questions. As is demonstrated by the rich range of ordered phases found in transition metal oxides, the low temperature physics of interacting lattice fermions can be quite complicated. The higher temperature physics is simpler, dominated by the crossover from a metal to a Mott insulator. Many believe that this crossover is a manifestation of a preempted quantum critical regime and should display universal physics analogous to that of the Bose-Hubbard model's normal fluid to Mott insulator crossover (see the discussion in Refs.[1,7]). Cold atoms experiments can be used to look for this scaling behavior: if one analyzes the density profiles with the procedure we have introduced and sees collapse into universal curves, this is compelling evidence that the crossover is governed by a quantum critical point. If the quantum critical regime persists to the same temperature scales as the Bose-Hubbard model, then current atomic experiments are already sufficiently cold to determine if the universal Mott-metal crossover scenario is correct.

Beyond determining if the Fermi system is critical, this analysis would also provide the dynamical critical exponent: a quantity whose value is currently unknown. Depending on the character of the metallic state, some filling-controlled Mott-metal transitions display $z = 2$, while others display $z = 4$[7]. Present experiments are capable of distinguishing between these scenarios.

One can explore an even broader range of open questions if in addition to measuring density profiles one also employs spectroscopic probes to measure dynamical response functions. For example, for lattice systems experimentalists have successfully measured modulation spectra[23], Bragg spectra[24], and radio-frequency spectra[25]. In the quantum critical regime, these response functions satisfy universal scaling forms $\chi(\omega, T, \Delta) = T^\eta \Psi_\chi(\Delta/T, \omega/T)$. In both fermionic and bosonic systems the details of these dynamic scaling functions are largely unknown.

Up to this point, all of our discussion has focused on what can be achieved with present experimental capabilities. As experiments reach lower temperatures, our analysis technique can be used to probe ever more fundamental physics. For example, one of the simplest symmetry breaking transitions of a Fermi liquid, that to a spin density wave, is ill-understood: beyond the Hertz-Millis theory, there exist an infinite number of marginally relevant coupling constants and its universality class is unclear[26,27]. Additionally, the competing instabilities seen in transition metal oxides have analogs in lattice fermion experiments, and offer an important and even richer set of open questions regarding quantum critical behavior and "avoided quantum criticality"[2–5,28]. Finally, the dynamics of the $O(2)$ model studied in the Bose-Hubbard model is relevant to dynamics near the Dirac point in graphene[13].

**Conclusions.** Despite the challenging technical requirements, current experiments are capable of using our protocol to study quantum criticality with cold atoms. Experiments on both lattice bosons and fermions have reached sufficiently low temperature[16], and boson density images display sufficiently high signal-to-noise, spatial resolution[18] and spectroscopic resolution[25].

**Supplementary Information** is linked to the online version of the paper at www.nature.com/nature.

**Acknowledgements**   We thank Mukund Vengalattore, Stefan Baur, Stefan Natu, Sourish Basu, Srivatsan Chakram, Eun-Ah Kim, Cheng Chin, Nate Gemelke, Ben Machta, Qi Zhou, Ed Taylor, Victor Gurarie, and Subir Sachdev for discussions. This work was supported by the National Science Foundation through Grant PHY-0758104 and by a grant from the Army Research Office with funding from the DARPA OLE program.

**Competing Interests**   The authors declare that they have no competing financial interests.

**Authors' contributions**   KRAH carried out all of the original research reported here. EJM provided advice and direction. KRAH and EJM contributed equally to writing the manuscript.

**Correspondence**   Correspondence and requests for materials should be addressed to Kaden R. A. Hazzard (email: kh279@cornell.edu).

**Supplementary information**

**Bose-Hubbard model.**   The Bose-Hubbard model[1] describes cold bosonic atoms in an optical lattice[2]. It is defined by the Hamiltonian

$$H = -t \sum_{\langle i,j \rangle} b_i^\dagger b_j + \sum_i \left[ \frac{U}{2} b_i^\dagger b_i^\dagger b_i b_i - \mu b_i^\dagger b_i \right] \quad \text{(S1)}$$

where $\sum_{\langle i,j \rangle}$ indicates a sum over nearest neighbors $i$ and $j$, and the operators $b_i$ and $b_i^\dagger$ annihilate and create bosons on site $i$. They satisfy the canonical commutation relation $[b_i, b_j^\dagger] = \delta_{ij}$. The parameters $t$ and $U$ are controlled by the depth $V_0$ of the optical lattice defined by $V(r) =$ $V_0(\cos(2\pi x/\ell)+\ldots)$. Energies are typically measured in terms of $E_R = \hbar^2\pi^2/(2m\ell^2)$. The values $t/U = 0.01, 0.0585$ shown in Fig. 3 correspond to lattice depths of $V_0/E_R = 18.1, 11.4$, respectively, for $^{87}$Rb in a lattice with spacing $\ell = 532$nm. This is found by numerically solving the non-interacting lattice problem and using the relations[2] $t = -\int d^3r\, w(\mathbf{r} - \mathbf{r_i}) \left[-\frac{\hbar^2}{2m}\nabla^2 + V(\mathbf{r})\right] w(\mathbf{r} - \mathbf{r_j})$ and $U = \frac{4\pi\hbar^2}{m} a \int d^3r\, |w(\mathbf{r})|^4$, where $w(\mathbf{r})$ is the Wannier state, and $\mathbf{r_i}, \mathbf{r_j}$ are the locations of neighboring sites.

**Finite temperature Gutzwiller theory.** We calculated the schematic finite temperature phase diagram in Fig. 1 within a finite temperature Gutzwiller approximation. Although only approximate this approach provides a qualitative picture of the role of temperature. More accurate phase diagrams can be obtained by using quantum Monte-Carlo techniques.

Following Fisher *et al.*[1] and Sachdev[1], we perform a Hubbard-Stratonovitch transform to decouple lattice sites: we introduce a new free field $\phi$ into the Bose-Hubbard Lagrangian and couple it to the $b$ operators by terms $-\sum_i (b_i^\dagger \phi_i^\dagger + \text{H.c.})$. The new fields' Lagrangian is chosen so that upon integrating out $\phi$ the original Bose-Hubbard Hamiltonian is reproduced. This field $\phi$ can be interpreted as the order parameter of the superfluid state. We formally integrate out the $b$ fields and expands the Lagrangian for the $\phi$ fields to quartic order in $\phi$ to obtain $\mathcal{L}_\phi = s|\phi|^2 + u/2|\phi|^4$. This procedure yields

$$s(T) = \frac{1}{zt} + \frac{1}{\mathcal{N}}\left[\sum_{m=0}^{\infty} \frac{m+1}{Um - \mu}\left(e^{-\beta(U/2)m(m-1)+\beta\mu m} - e^{-\beta(U/2)m(m+1)+\beta\mu(m+1)}\right)\right] \quad \text{(S2)}$$

with $\mathcal{N} \equiv \sum_{m=0}^{\infty} e^{-\beta((U/2)m(m-1)-\mu m)}$. The sums converge rapidly. The phase boundary is given by setting $s(T) = 0$. At $T = 0$ this reproduces the standard Gutzwiller theory.

**Non-universal contributions.** In order to see the collapse described in the main text, one needs to subtract off the non-universal contributions to the density ($n_0$) and compressibility ($\kappa_0$). The non-universal contribution to any on-site observable is obtained by evaluating that observable at $t = 0$ and at the $U$, $T$, and $\mu$ of interest. This non-universal part may in principle be measured experimentally by increasing the depth of the optical lattice. In practice, there are a number of technical hurdles: equilibration is difficult in very deep lattices[3], as is control and measurement of temperature[4]. Additionally, for very deep lattices the Bose-Hubbard description itself breaks down[5–12,25]. Alternatively, it is straightforward to analytically calculate the non-universal contribution, as it reduces to a single site problem. For example, for the Bose-Hubbard model's density one finds

$$n_0 = \frac{\sum_{n=0}^{\infty} n e^{-\beta \epsilon_n}}{\sum_{n=0}^{\infty} e^{-\beta \epsilon_n}} \tag{S3}$$

with

$$\epsilon_n = \frac{U}{2} n(n-1) - \mu n. \tag{S4}$$

These sums converge quickly with $n$, and typically only a few terms are needed. Note that here, as occasionally throughout the manuscript, we set $k_B = \hbar = 1$. The subtraction of the non-universal contribution requires knowledge of the central chemical potential (as may be obtained by fitting to a regime where the equation of state is known, for example the cloud wings or at high temperatures).

**Time of flight expansion.** As an alternative to density profiles, one can study quantum criticality through time-of-flight expansion in which the trapping potential and interactions are turned off and the cloud is allowed to expand. At long times this maps the momentum distribution of the particles to the real space distribution. At low momenta, the system's behavior is fully universal, with no need to subtract a non-universal contribution. Consequently one does not require knowledge of the central chemical potential. Kato *et al.*[13] show a representative calculated image of the momentum distribution in a quantum critical regime of the Bose-Hubbard model.

In typical cold atom experiments the inhomogeneous broadening from the trap makes it very difficult to extract quantum critical signatures from the expansion images: multiple regions of the trap contribute to the observed momentum distribution, including those far from the quantum critical regime. This difficulty can be circumvented by engineering a flat bottomed trapping potential. Such flat traps may also be advantageous for density probes of quantum criticality.

**Finite density $O(2)$ model.** This section describes more precisely the crossover between the dilute Bose gas and the $O(2)$ universality classes, and outlines open questions. Figure 4 illustrates the regions governed by each universality class. Along the Mott lobe edges, the physics is that of the dilute Bose gas (Figure 4a, shaded blue regions). On the large-$\mu$ side of the Mott lobe, the relevant excitations are a dilute Bose gas of particles, while on the small-$\mu$ side, they are holes. Near the line of particle-hole symmetry passing through the tip of the lobe, both particles and holes are equally important and the physics is in the $O(2)$ universality class (Figure 4a, shaded orange region). Both of these universal structures are captured by the "finite density $O(2)$" model with imaginary time action $S = \int d^d r d\tau \mathcal{L}(\mathbf{r}, \tau)$ defined by the Lagrangian

$$\mathcal{L} = -\phi^* \left[ (\partial_\tau - \mu)^2 - c^2 \nabla^2 + s \right] \phi + \frac{g}{2} |\phi|^4 \tag{S5}$$

where $\phi$ is a complex bosonic field, evaluated at $\mathbf{r}, \tau$ in this expression. Here $\mu$ is the chemical potential, which controls the relative energy cost of holes and particles, $s$ is the tuning parameter for the $\mu = 0$ phase transition, $c$ the excitation velocity when $\mu = 0$, and $g$ is the effective interaction strength. When $\mu$ is non-zero, at sufficiently low energies this reduces to the dilute Bose gas Lagrangian $\mathcal{L}_{\text{DBG}} = \psi^* \left[ \partial_\tau - \mu - \nabla^2/(2m) \right] \psi + (g/2) |\psi|^4$ (with $\psi$ a simple rescaling of $\phi$) since the quadratic time derivative is *irrelevant* in the renormalization group sense. When $\mu = 0$, this reduces to the $O(2)$ model defined by the Lagrangian $\mathcal{L}_{O(2)} = \phi^* \left( -\partial_\tau^2 - c^2 \nabla^2 + s \right) \phi + (g/2) |\phi|^4$. The model defined by Eq. (S5) predicts a scaling function[9, 14, 15]

$$n_u(\Delta_+, \Delta_-, T) = T^d \Psi_{O(2)+\mu} \left( \frac{\Delta_+}{T}, \frac{\Delta_-}{T} \right) \tag{S6}$$

where $\Delta_\pm$ are the relevant gaps/superfluid stiffnesses at $T = 0$. This scaling function is universal and reduces to the dilute Bose gas and $O(2)$ scaling in the appropriate limits. [To see this in the dilute Bose gas case, note that if for large $\Delta_-/T$ the scaling function goes to $\Psi(\Delta_+/T, \Delta_-/T) \to (\Delta_-/T)^{d/2} \Psi_r(\Delta_+/T)$ then $n_u \to T^{d/2} \Delta_-^{d/2} \Psi_r(\Delta_+/T)$. This reproduce the $z = 2$ scaling expected for the resulting dilute Bose gas case if $\Psi_n(\Delta_+/T) = \Delta_-^{d/2} \Psi_r(\Delta_+/T)$ is identified as the universal dilute Bose gas scaling function introduced in the main text.] This more general finite density $O(2)$ universal physics describes the entire the shaded green region of Fig. 4a.

A particularly insightful way to view the finite density $O(2)$ crossovers is illustrated in Fig. 4b. This corresponds to a slice through Fig. 4a in the region described by Equation (S6).

The low temperature behavior is dominated by the two dilute Bose gas quantum critical points, and the $O(2)$ physics emerges at higher temperatures where the two fans overlap. Not only does the scaling in Eq. (S6) describe the physics in these fans, it also describes the low temperature phases.

**Calculating density profiles of one-dimensional hardcore bosons.** We calculate our density profiles by mapping the one-dimensional, hardcore ($U \to \infty$), trapped lattice bosons described by the Hamiltonian

$$H = -t \sum_i \left[ b_i^\dagger b_{i+1} + \text{H.c.} + \frac{U}{2} b_i^\dagger b_i^\dagger b_i b_i + (V_i - \mu) b_i^\dagger b_i \right], \tag{S7}$$

with harmonic trapping potential $V_i$, onto non-interacting fermions by the Jordan-Wigner transformation[16]

$$f_i \equiv \left[ \prod_{j<i} (1 - 2 b_j^\dagger b_j) \right] b_i. \tag{S8}$$

Note that $f_i$ and $f_i^\dagger$ satisfy the canonical anticommutation relation for fermions, $\{f_i, f_j^\dagger\} = \delta_{ij}$. This gives the non-interacting Fermi Hamiltonian

$$H = -t \sum_i \left[ f_{i+1}^\dagger f_i + \text{H.c.} + (V_i - \mu) f_i^\dagger f_i \right]. \tag{S9}$$

We numerically find the single particle eigenstates $\phi_i^{(\alpha)}$ with energy $E_\alpha$. The bosonic density at site $i$ is then equal to the fermionic density at site $i$ by Eq. (S8), and thus

$$n_i = \sum_\alpha \frac{1}{e^{\beta E_\alpha} + 1} |\phi_i^{(\alpha)}|^2. \tag{S10}$$

**Other cold atoms systems displaying quantum criticality.** For reference, we provide a partial list of other quantum phase transitions in cold atoms systems. The most experimentally mature systems are: magnetic transitions in spinor gases[17], nematic transitions in dipolar gases[18], superfluid and magnetic transitions in partially polarized resonant Fermi gases[19], and transitions from fully polarized or fully paired phases to a partially polarized (FFLO) phase in one dimensional clouds of fermions[20]. There are also potentially a large variety transitions between of magnetic phases in multicomponent gases and mixtures[6]. Finally, there are ample opportunities to study the *trivial* dilute gas to vacuum transitions, governed by the chemical potential tuned zero temperature phase transition from a state with non-zero to zero density. This physics is found near the edge of every atomic cloud. Although well understood, these latter transitions are a good test of the analysis techniques.

Universal scaling behavior has already been experimentally studied in the two-dimensional dilute trapped Bose gas using techniques related to the ones we discuss here[21]. In the two dimensional Bose system which they were looking at, dimensional analysis alone suffices to provide the collapse: all of the irrelevant couplings are zero in the bare Hamiltonian describing this system.

**Quantum Monte Carlo parameters.** We calculate the densities for the two-dimensional square lattice Bose-Hubbard model using worm algorithm quantum Monte Carlo algorithm[22] as implemented in the ALPS simulation package[20]. We performed a sufficient number of equilibration sweeps (10,000) and evaluation sweeps (30,000) to obtain accurate estimates of the density. Typical stochastic errors in the density were $\sim 0.1\%$, but these are amplified when we take derivatives to extract the compressibility. This stochastic error is comparable to imaging noise of radially averaged density profiles in real experiments[18]. As seen in Fig. 3, our approach is robust against such noise. Over most of the parameter space simulated, we find that systematic errors from the finite equilibration time are significantly smaller than the stochastic errors. We explore possible systematic errors using two methods: (1) running with longer equilibration times and (2) a jacknife binning analysis. The simulations were carried out for a system size of $10 \times 10$ lattice sites. Our results were insensitive to the finite size effects, except in the *classical* critical regime, mimicking the effects of a real trapping potential.

**Figure 1   Quantum critical crossovers in the Bose-Hubbard model. a**, Mott insulator to superfluid phase boundaries for (left to right) temperatures $T/U = 0.00, 0.06, 0.12, \ldots, 0.96$, calculated via finite temperature Gutzwiller theory (see Supplementary Information) in $d = 2$. The parameters $t$, $\mu$, and $U$ are the tunneling rate, chemical potential, and on-site interaction energy of the Bose-Hubbard model. Paths (1,2) are governed by the dilute Bose gas universality class, and path (3) (passing through the tip) is governed by the $O(2)$ universality class. See Supplementary Information for more details. **b**, Slice through the quantum phase transition at fixed $t/U$. Slices with fixed $\mu/U$ are similar. The abbreviations "SF," "MI," and "QC" denote the superfluid, Mott insulator-like normal fluid, and quantum critical regime, with the Mott insulator being strictly defined only at $T = 0$. The shaded region indicates where the physics is governed by the quantum critical point. The dashed lines represent smooth crossovers between the qualitatively distinct "QC" and "low temperature" regions. Deep in the former, $T$ is the only relevant energy scale.

**Figure 2   Extracting universal behavior and dynamical critical exponents from density profiles of the trapped one-dimensional Bose-Hubbard model. a**, Exact density profiles of the one dimensional harmonically trapped hard core Bose-Hubbard model for $N = 70$ particles at temperatures $\bar{T} \equiv T/t = 0.1, 0.24, 0.38$, with larger temperatures corresponding to lower central density. Here, $r$ is the radial displacement in the trap and $d$ is the lattice spacing. These density profiles are non-universal: for example, they depend on temperature. **b**, Our construction for obtaining universal scaling curves applied to this system, plotting $\bar{\kappa}\bar{T}^{d/z}$ versus $\bar{n}/\bar{T}^{1/2}$ (defining $\bar{\kappa} \equiv \kappa - \kappa_0$, $\bar{n} \equiv n - n_0$, and $\bar{T} = T/t$) for this

$d = 1$, $z = 2$ transition and temperatures $\bar{T} = 0.1, 0.17, 0.24, 0.31, 0.38$. The compressibility is approximated by $\kappa = \partial n/\partial \mu \approx (\partial n/\partial r)/(m\omega^2 r)$, where $\omega$ is the trap frequency, and $\partial n/\partial r$ is obtained by numerically differentiating the density. Lower temperatures display a larger region of collapse. We observe good collapse up to $T \sim 0.25t$, and that the analysis accurately reproduces the homogeneous infinite system's scaling curve (shaded gray line) within $\lesssim 10\%$ for $T \gtrsim 0.05t$ for the transition near $n = 1$ ($\bar{n} < 0$) and for $T \gtrsim 0.15t$ for the transition near $n = 0$ ($\bar{n} > 0$). This collapse occurs even for drastically different density profiles obtained by adjusting the trap depth in place of temperature (not shown). With moderately larger particle numbers ($N \sim 200$, not shown), the simulated data at low temperatures even more accurately reproduces the infinite homogeneous system's universal scaling function (the extracted curve lies within the shaded gray region).

**Figure 3 Extracting universal behavior and dynamical critical exponents from density profiles of the trapped two-dimensional Bose-Hubbard model.** Each panel shows an application of our analysis procedure to simulated density profiles (as described in text) for temperature $T/t = 1/4, 1/2, 1, 2$, and $4$ using the data in a density range $|n - 1| < 0.15$. The symbols $\bar{n}$, $\bar{\kappa}$, and $\bar{T}$ are defined in Fig. 2. Lower temperature curves are identified by noting that they span a wider $\bar{n}$ range. On the left (**a** and **c**), we plot $\bar{\kappa}$ versus $\bar{n}\bar{T}^{-1}$, while on the right (**c** and **d**) we plot $\bar{\kappa}\bar{T}^{-1}$ versus $\bar{n}\bar{T}^{-2}$: these will show collapse if the dynamical critical exponent is respectively $z = 2, 1$. Top (**a** and **b**) shows data with $t/U = 0.0400$, which should be described by the dilute Bose gas universality class ($z = 2$). Bottom (**c**

and **d**) shows data with $t/U = 0.0585$, which should be better described by the $O(2)$ rotor model universality class ($z = 2$): the tip of the $n = 1$ Mott lobe in the homogeneous system is at $t/U = 0.0593$. As expected, we see collapse in **a** and **d**. The scatter in data points corresponds to stochastic noise in our Monte-Carlo simulations amplified by the differentiation. This noise is of comparable size to what would be seen in an experiment.

**Figure 4   (S1) Universality classes of the Bose-Hubbard model. a**, Annotated zero temperature Bose-Hubbard model phase diagram (cf. Fig. 1). Transitions via paths (1-2) are described by the dilute Bose gas (DBG) universality class while path (3) is described by the $O(2)$ universality class. Shaded regions schematically depict where each universality class holds: DBG physics governs the blue region along the lobe edge and $O(2)$ rotor physics governs the orange "bowtie" shaped region near the tip. The entire green region is described by the finite density $O(2)$ model. **b**, Finite temperature phase diagram along path (4). At zero temperature, this path crosses two quantum phase transitions from a superfluid to a Mott insulator, and back. Each phase transition is in the DBG universality class and displays a quantum critical fan, inside of which the temperature sets the only length scale. $O(2)$ physics is recovered in the particle-hole symmetric regions where the quantum critical fans overlap. The "finite density $O(2)$" model universally governs this entire phase diagram including all of the crossovers (dashed lines) and phase transitions (solid lines) shown in this panel.

**a** 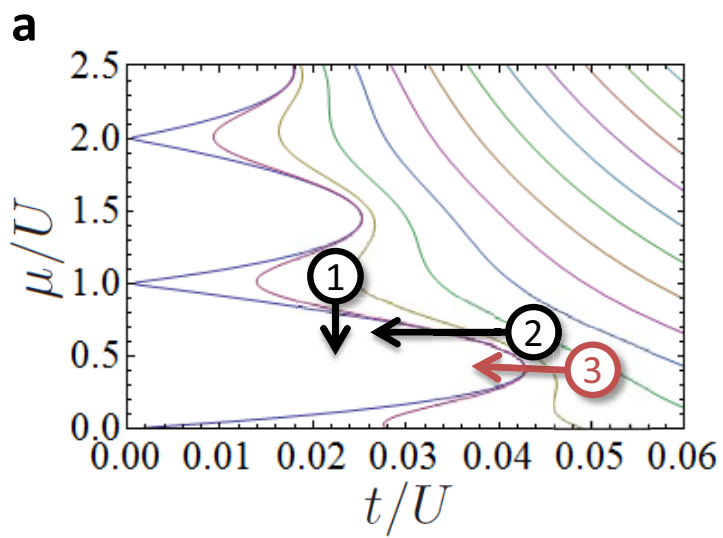

**b** 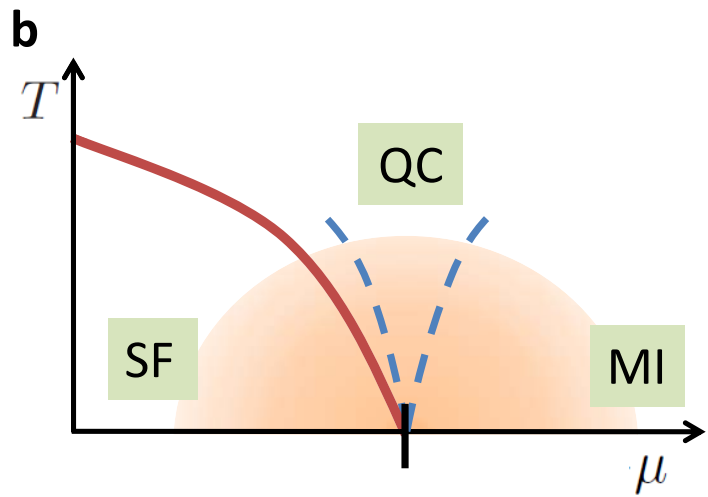

Authors: Kaden R. A. Hazzard and Erich J. Mueller
Figure 1

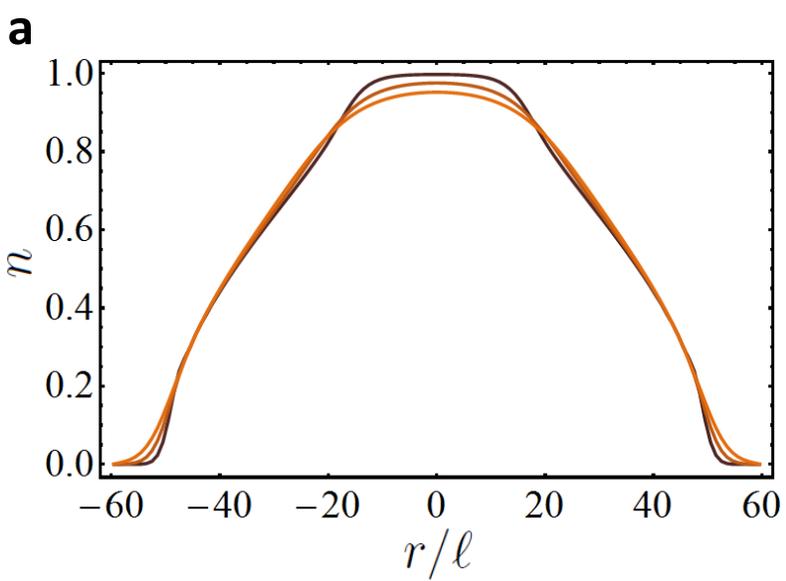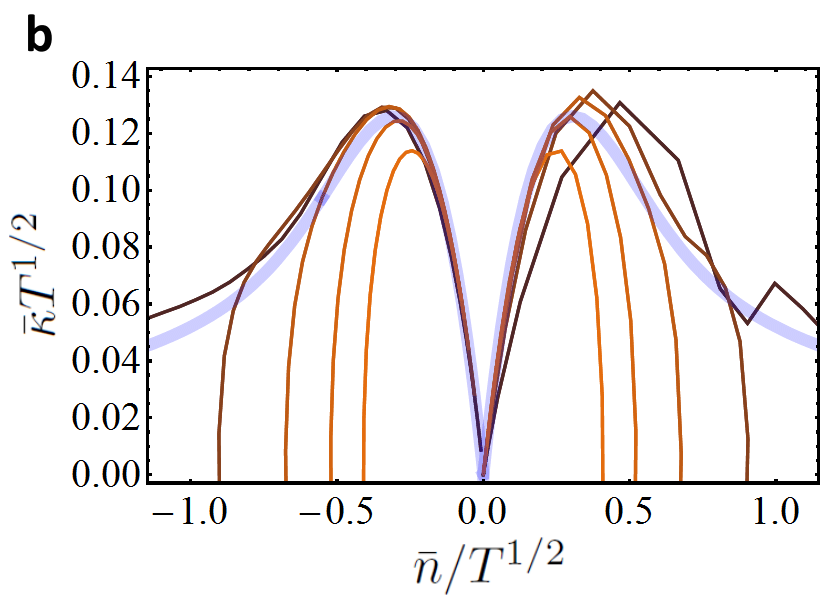

Authors: Kaden R. A. Hazzard and Erich J. Mueller
Figure 2

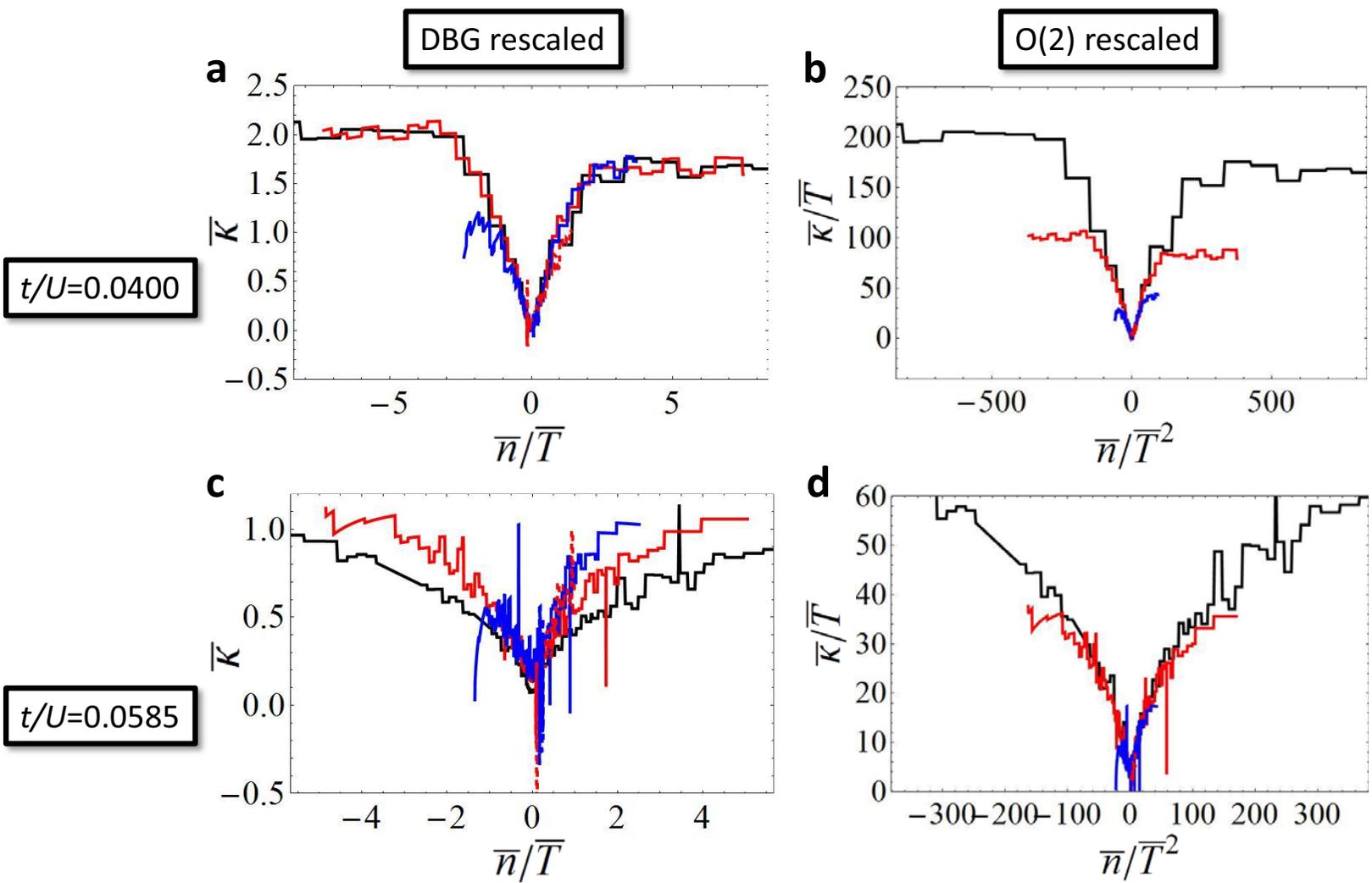

Authors: Kaden R. A. Hazzard and Erich J. Mueller
Figure 3

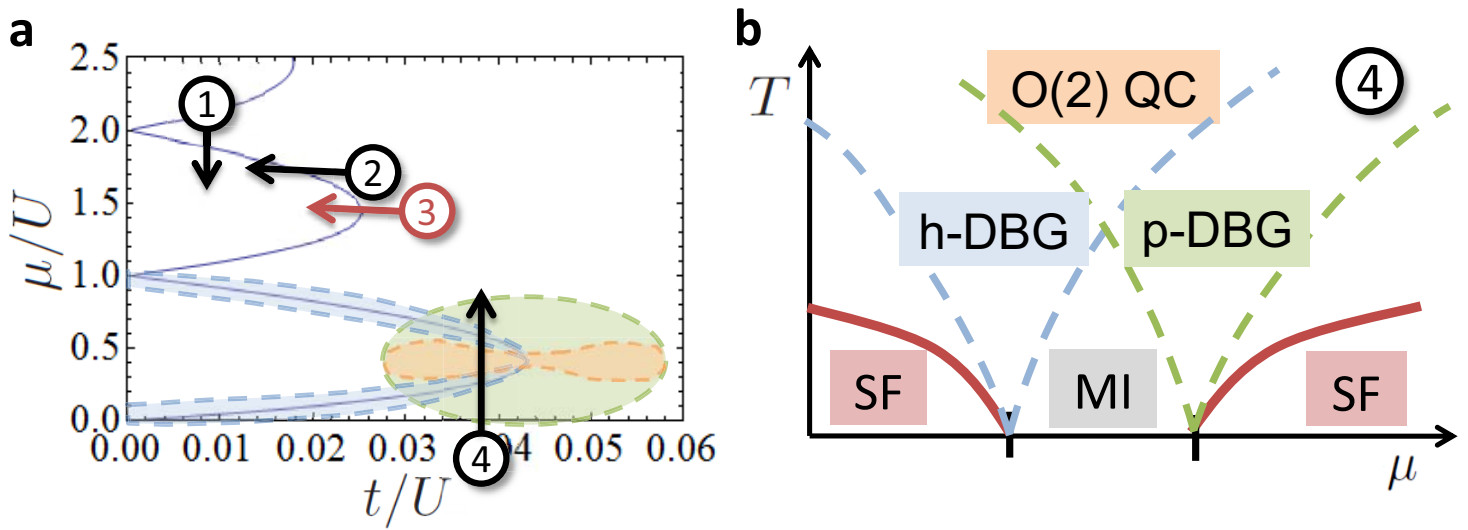

Authors: Kaden R. A. Hazzard and Erich J. Mueller
Figure 4 (supplementary info Figure S1)